\documentclass[aps]{revtex4}
\usepackage{graphicx}

\begin{document}

\title{Josephson effect through a quantum dot array}
\author{F.S. Bergeret, A. Levy Yeyati, A. Mart\'{i}n-Rodero}
\affiliation{Departamento de F\'{\i}sica Te\'{o}rica de la Materia Condensada
C-V, Universidad Aut\'{o}noma de Madrid, E-28049 Madrid, Spain}

\begin{abstract}
We analyze the ground state properties of an array of quantum dots connected in series between superconducting electrodes. This system is represented by a finite Hubbard chain coupled at both ends to BCS superconductors. The ground state is obtained using the Lanczos algorithm within a low energy theory in which the bulk superconductors are replaced by effective local pairing potentials. We study the conditions for the inversion of the sign of the Josephson coupling ($\pi$-junction behavior) as a function of the model parameters. Results are presented in the form of phase diagrams which provide a direct overall view of the general trends as the size of the system is increased, exhibiting a strong even-odd effect. The analysis of the spin-spin correlation functions and local charges give further insight into the nature of the ground state  and how it is transformed by the presence of superconductivity in the leads. Finally we study the scaling of the Josephson current with the system size and relate these results with previous calculations of Josephson transport through a Luttinger liquid.            
\end{abstract}

\maketitle

\section{Introduction}
 Advances in the fabrication of hybrid nanostructures consisting  of  quantum dots, molecules or carbon nanotubes  attached to  conducting electrodes have lead to an increasing  interest  in  the electronic and transport properties through such low dimensional devices. In most of these systems electronic correlations play an important role leading for instance to the Kondo effect \cite{cronenwett,buitelaar,ralph}. In particular, when the leads are superconductors  a natural question that arises is whether a supercurrent may flow through the structure. The interplay between Josephson and correlation effects have received considerable attention in the last years \cite{glazman,arovas,ambegaokar,us,egger,choi,tmartin,kroha,vecino,hewson}. A paradigmatic example of this interplay is the prediction of a transition to a $\pi$-junction behavior as a function of some relevant system parameters \cite{spivak}. This prediction has been recently confirmed experimentally both in carbon nanotubes and semiconducting nanowires \cite{bouchiat,leo}. In spite of all the recent theoretical activity in this field there are still many issues to be explored. While most theoretical works have concentrated in analyzing the Josephson effect through a single level Anderson model coupled to superconducting electrodes, the analysis of more complex models taking into account the spatial extension of the correlated region is still lacking. In this respect some works addressing the double quantum dot situation have appeared recently \cite{us,lopez}. A natural extension of the Anderson model to describe a correlated region with a finite size is provided by the Hubbard model. Such a model could describe different physical situations like {quantum dot arrays, molecular junctions or  nanowires}. In the present work we will take the linear Hubbard chain as a prototypical example. Some aspects of this model have been recently discussed in Ref. \cite{scalapino}.  On the other hand, studies of Luttinger liquids in contact with superconductors \cite{affleck,caux} provide predictions on the Josephson effect which the model should obey in certain limits. As we  show in this work  the superconductor-chain-superconductor  system reveals several novel correlation effects which are not found in the strictly zero dimensional Anderson model and which are associated with the finite spatial {extension of the system.}

The paper is organized as follows: In section II we introduce the model and discuss the approximation scheme used to obtain the ground state properties of the system. {We will argue that these properties can be   adequately described}  by a low energy theory in which the bulk superconducting electrodes are replaced by local pairing potentials.{ In order to test the validity of this approximation we consider  in section III  the single site case (single level Anderson model) and compare the results with the ones obtained using other methods like Hartree-Fock and finite U slave boson mean field. In section IV we present the phase diagrams illustrating the regions in parameter space where $\pi$-junction behavior is found for different number  of sites in the array.} Section V is devoted to analyze the spin-spin correlations and the spatial distribution of the spin excess which characterize the $\pi$-junction state. {Such analysis  is useful for the full characterization of  the interplay between Kondo, antiferromagnetic and pairing correlations}. Finally in section VI we present results for the current-phase relation and its scaling behavior as the number of sites is increased. By means of these results we discuss the connection with previous studies for Luttinger liquids connected to superconducting leads \cite{affleck,caux}.   The paper is closed with some concluding remarks.

\section{Model and method}
In the present work we shall assume that the system can be described by the following Hamiltonian 
\begin{equation}
\hat{H}=\hat{H}_L+\hat{H}_R+\hat{H}_{chain}+\hat{H}_t	
\label{hamiltonian1}
\end{equation}
where $\hat{H}_{L(R)}$ corresponds to the left (right) superconducting lead, characterized by order parameters $\Delta_{L(R)}exp(i\phi_{L(R)})$ and 
$\hat{H}_{chain}$ denotes the finite Hubbard chain Hamiltonian with $N$ sites

\begin{equation}	\hat{H}_{chain}=t\sum_{i=1}^N(\hat{c}_{i\sigma}^{\dagger}\hat{c}_{i+1\sigma}+h.c.)+U\sum_{i=1}^N\hat{n}_{i\uparrow}\hat{n}_{i\downarrow}+\epsilon\sum_{i=1}^N\hat{n}_{i\sigma}
\label{hubb}
\end{equation}
where the field operators $\hat{c}_{i\sigma}^{\dagger}$ creates a particle in site $i$ with spin $\sigma$, $t$ is the hopping parameter  within the chain, $U$ is the on site electron interaction and $\epsilon$ is the site energy. 
The coupling between the chain and the leads is described by the term $\hat{H}_t$.
The determination of the ground state properties of this model is a formidable task which requires some approximation scheme. Several theoretical approaches have been used in the single site case ($N=1$), including numerical renormalization group \cite{choi}, quantum Monte Carlo \cite{egger}, finite order perturbation theory \cite{vecino}, Hartree-Fock \cite{arovas} or slave-boson mean-field \cite{us,lopez}.

A great simplification can be made by integrating out the electronic degrees of freedom of the superconducting leads in the way it was proposed in Ref. \cite{affleck}. This procedure leads to an effective low energy theory in which each superconductor is replaced by a single site with an effective pairing potential $\tilde{\Delta}_{L(R)}$. Also the hopping terms between the leads and the chain are replaced by  effective parameters $\tilde{t}_{L(R)}$. Thus, the terms $\hat{H}_{L(R)}$ and $\hat{H}_t$ take the form
\begin{eqnarray}
\hat{H}_{L(R)}&=&\tilde{\Delta}_{L(R)}e^{i\varphi_{L(R)}}\hat{c}_{L\uparrow}^{\dagger}\hat{c}_{L\downarrow}^{\dagger}+ h. c.\nonumber\\
\hat{H}_t&=&(\tilde{t}_L\hat{c}_{L\sigma}^{\dagger}\hat{c}_{1\sigma}+H.c.)+(\tilde{t}_R\hat{c}_{N\sigma}^{\dagger}\hat{c}_{R\sigma}+h.c.)
\label{hlrht}
\end{eqnarray}
We shall refer to this simplified description of the leads as the zero bandwidth model (ZBWM).
In principle the effective parameters in this approach have to be determined from the bare parameters by means of a self-consistency condition. As discussed by Affleck {\it et al.}  \cite{affleck} using  a renormalization group  analysis the effective parameters, with positive or negative  interactions,  flow to the condition of zero or perfect Andreev reflection respectively in the limit of an infinite chain. However, for the finite chain case these arguments are no longer valid. We shall adopt here the simplifying assumption that $\tilde{\Delta}_{L,R}=\tilde{t}_{L,R}=t$ without an attempt to fine tune them within the range of parameters considered. This is a reasonable choice as far as one is interested in the qualitative trends rather than in the detailed quantitative results. As we will show in the next section, taking the one site case as a test case, the ZBWM yields a phase diagram in  good agreement with known results from other approaches that take into account the full spectrum of the superconducting leads. 
One should mention that the Hamiltonian in the ZBWM couples in principle $4^{(N+2)}$ configurations corresponding to the grand canonical ensemble. This number can be substantially reduced by projecting into subspaces with well defined spin. In order to reach the maximum possible system size we use the Lanczos algorithm for the determination of the system ground state \cite{dagottoRMP}. This allows us to obtain results for systems up to 10 sites in the chain.   

\section{The single level Anderson model as a test case}
In this section we shall consider the single level case in order to test the validity of the ZBWM for the superconducting leads. Previous studies have shown that this system undergoes a quantum phase transition to a $\pi$-state around the region in parameter space corresponding to the Kondo regime for  normal electrodes \cite{arovas,egger,choi,vecino}. This state is characterized by the presence of an unscreened spin $1/2$ in the dot and by a reversal of the sign of the Josephson current-phase relation. This is associated with a minimum in the ground state energy at $\phi=\phi_L-\phi_R=\pi$. Outside the Kondo region, the ground state of the system for superconducting leads is non-degenerate with a minimum in the energy at $\phi=0$. This state is usually referred to as the $0$ phase. The transition between the $0$ and the $\pi$ phases takes place through a sequence of two intermediate phases called $0'$ and $\pi'$ where both minima are present \cite{arovas}.

\begin{figure}[t]
\includegraphics[scale=0.8]{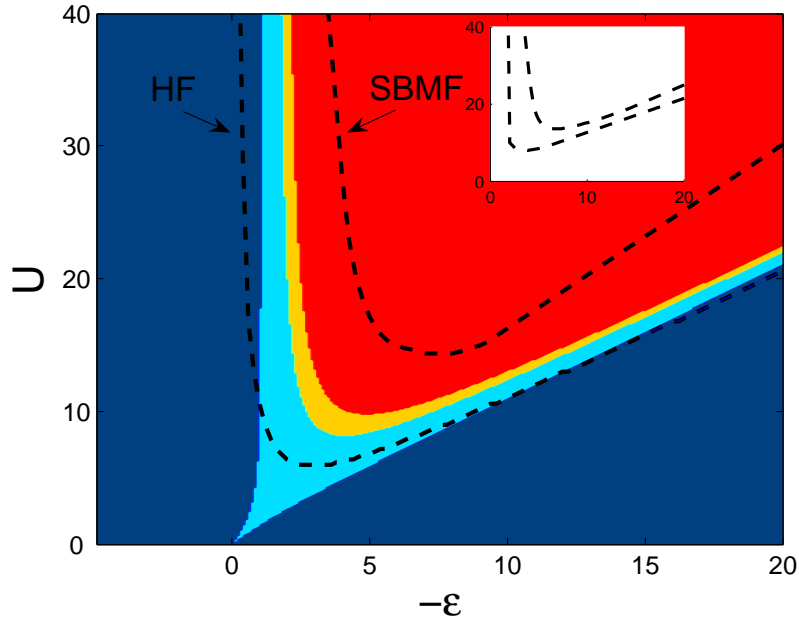}
\caption{(Color online) $\epsilon-U$ phase diagram of the single quantum dot case in the ZBWM. 0, 0' $\pi'$ and $\pi$ regions are indicated by blue, light blue, yellow and red respectively. Dashed lines indicate the boundary of the $\pi$ region when calculated using the HF and the SBMF approximations. The inset show these boundaries when taking into account the full spectrum of the superconducting electrodes with $\Gamma=t$. The parameters are taken in units of $t=\Delta$}. 
\label{diag1}
\end{figure}

The exact phase diagram in the $\epsilon-U$ space obtained in the ZBWM is shown in Fig. \ref{diag1}. For comparison we have also calculated the phase diagram in the ZBWM  using both the Hartee-Fock (HF) and the finite $U$ slave boson mean field (SBMF) approximations \cite{ruckenstein,us}. The dashed lines in Fig. \ref{diag1} indicate the boundary of the $\pi$-region within these two approximations. One can notice that the HF approximation tends to overestimate the stability of the pi-phase for lower $U$ values while the SBMF tends to do the opposite. On the other hand both approaches yield phase diagrams which are in qualitative agreement with the exact result in the ZBWM.

The fact that the exact phase boundaries in the ZBWM  lie in between this two different mean field approximations allows one to extrapolate some conclusions to the case of superconducting electrodes with their full spectrum. Although the full model have been studied using more accurate methods like numerical renormalization group \cite{choi} and quantum Monte Carlo \cite{egger}, a description of the complete phase diagram using these type of methods is not available. However, the results for the full model that can be readily obtained using HF and SBMF yield a rather precise idea of the form of this diagram. In fact the results exhibited in the inset in Fig. \ref{diag1} for both approximations show that  for the full model the diagram is quite similar to the the one obtained within the ZBWM  when making the correspondence $t \rightarrow \Gamma$, where $\Gamma$ is the standard tunneling rate to the leads. As in the discrete case one would expect that the boundaries of the exact diagram should lie somewhere in between. We therefore conclude that the ZBWM  with the prescription $t \rightarrow \Gamma$ provides a good qualitative description of the ground state properties in this kind of systems.

Finally, a word of caution should be said regarding the applicability of the ZBWM to the case of normal leads. While in the superconducting case the ground state properties are rather insensitive to the detailed structure of the leads,  this is no longer the case when the superconducting gap goes to  zero and the leads become normal. One can nevertheless rely on the ZBWM for a  rough qualitative description of the system in the normal state \cite{allub}.   
\begin{figure}[t]
\includegraphics[scale=0.8]{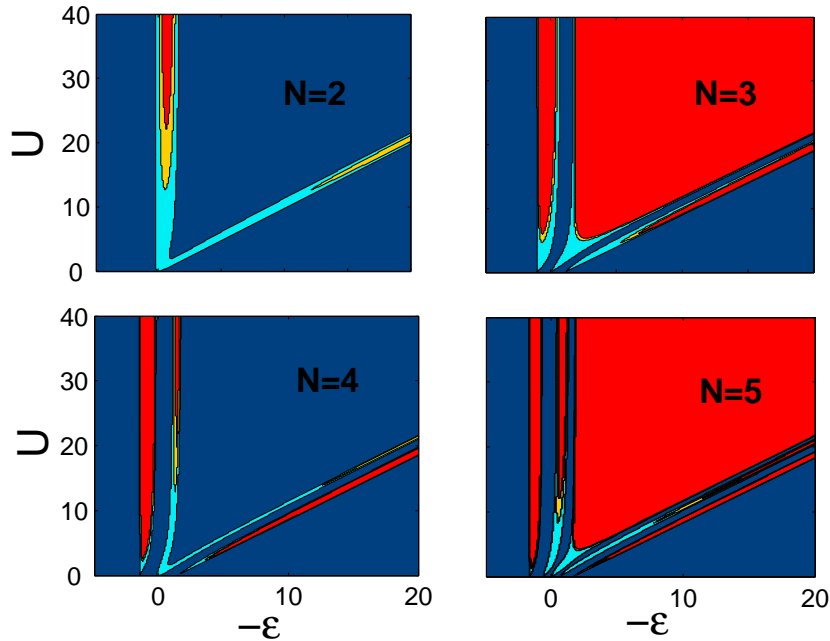}
\caption{(Color online)  $\epsilon-U$ phase diagrams for $N=2,3,4,5$. The color convention is the same as in Fig1.}
\label{diagrams}
\end{figure}

\section{Phase diagrams for the multi-dot case}

In what follows we analyze the evolution of the ground state properties of systems containing an increasing number of sites in the Hubbard chain and within the  ZBWM for describing the leads.
The overall properties of these systems are best illustrated by the evolution of the corresponding $\epsilon-U$ phase diagrams with increasing $N$. In Fig. \ref{diagrams} we show these diagrams for $N=2,3,4,5$.
As can be observed all these diagrams keep a certain resemblance with the single dot one (Fig. \ref{diag1}) in the sense that there appear lines separating the zero regions for the empty ($\epsilon>0$) and fully occupied ($\epsilon_0+U<0$) dots limit. However, one can notice a strong even-odd effect which manifests in the presence or absence of a $\pi$ region around the line $\epsilon\approx-U/2$ corresponding to the half-filled case. Moreover,  a distinct feature is the appearance of new branches of $\pi$ regions with increasing $N$ around the lines $\epsilon=0$ and $\epsilon+U=0$.
For even $N$ the absence of a $\pi$-phase around the half-filled region is explained by the fact that in this case there are $N$ electrons in the dots region which couple antiferromagnetically forming a singlet state with total spin $S=0$. On the contrary, for an odd number of sites at half filling there is always an unpaired electron which leads to a ground state with $S=1/2$, corresponding to the $\pi$-state. 

\begin{figure}
\includegraphics[scale=1]{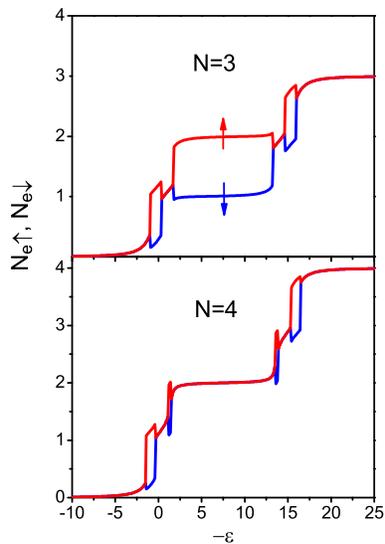}
\caption{(Color online) Total charge per spin in the dots region for $U=15$ for $N=3$ (upper panel) and $N=4$ (lower panel). }
\label{charge}
\end{figure}

The other relevant feature, namely the appearance of new branches of $\pi$-regions around the lines $\epsilon=0$ and $\epsilon+U=0$ can be readily associated with the condition of a total charge in the chain region corresponding to an odd integer. This can be checked by analyzing the behavior of the total mean charge per spin ($N_{e\uparrow},N_{e\downarrow}$) of  chains with $N=3,4$ as a function of the level position for a given value of $U$ (Fig. \ref{charge}). As can be observed the $\pi$-regions are associated with the magnetic symmetry breaking ($N_{e\uparrow}\neq N_{e\downarrow}$) which occurs around the regions where the total mean charge in the chain is an odd integer.
\section{Spin-spin correlations and spatial distribution of spin excess}
\begin{figure}
\includegraphics[scale=1]{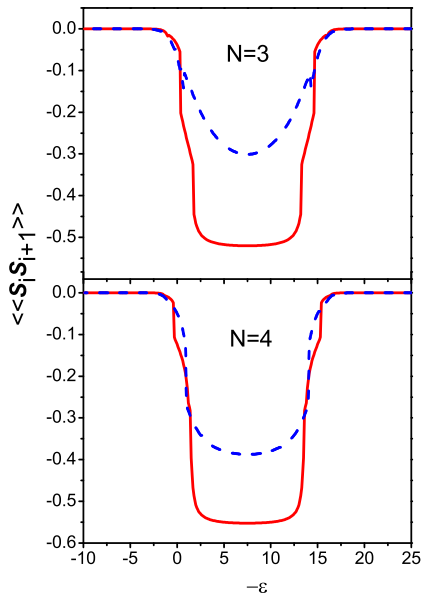}
\caption{(Color online) Average first-neighbors spin-spin correlation inside the chain for the  same parameters as in Fig. \ref{charge}. The red solid (blue dashed) line corresponds to the superconducting (normal) case.}
\label{sisj}
\end{figure}

\begin{figure}[t]
\includegraphics[scale=1]{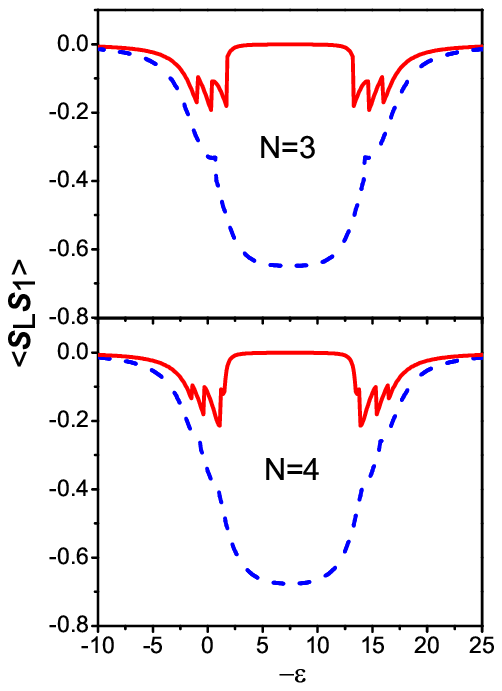}
\caption{(Color online) Spin correlation between the left lead and the first site of the chain. Same parameters and color convention as in Fig. \ref{sisj}.}
\label{sls1}
\end{figure}

An additional insight into the nature of the correlated ground state can be obtained from the analysis of spin correlations. For this purpose we calculate both the intra chain spin correlations ($<\textbf{\textit{S}}_i\textbf{\textit{S}}_j>$, $i,j=1,N$) and the correlation between spins in the chain and one of the leads $<\textbf{\textit{S}}_{L}\textbf{\textit{S}}_i>$.  
In Fig. \ref{sisj} we show the average correlation between neighboring sites inside the chain, $ <<\textbf{\textit{S}}_i\textbf{\textit{S}}_{i+1}>>\equiv\sum_{i=1}^{N-1} <\textbf{\textit{S}}_i\textbf{\textit{S}}_{i+1}>/(N-1)$ for $N=3,4$ both in the normal and in the superconducting state as a function of the level position for a fixed value of $U$. As could be expected this correlation acquires an antiferromagnetic character reaching its maximum (negative) value around the condition of half-filling. On the other hand one can clearly notice that the effect of the superconducting electrodes is to increase the antiferromagnetic correlations with respect to the normal state \cite{us}. 
The correlation between the spins in the leads and in the chain is revealed by analyzing the quantity $<\textbf{\textit{S}}_{L}\textbf{\textit{S}}_1>$ which is shown in Fig. \ref{sls1} for the same set of parameters as in the previous figure. Again in this case these correlations have an antiferromagnetic character reaching in the normal state their maximum negative value around the condition of half-filling. However, in contrast with the internal spin correlations, they are strongly suppressed around half-filling for superconducting electrodes due to the competing effect of pairing correlations.

\begin{figure}[t]
\includegraphics[scale=0.8]{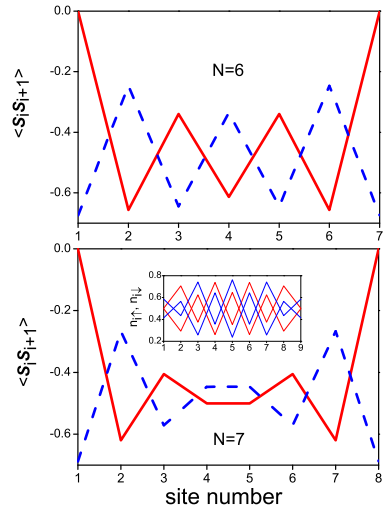}
\caption{(Color online) Spatial distribution of first neighbors spin correlation in the half-filled case for $U=20$. The inset in the lower panel represents the charge per spin on each site of the system. Site number 1 indicates left electrode. We use the color convention of Fig. \ref{sisj}.}
\label{cargacorrel}
\end{figure}

It is also instructive to analyze the spatial distribution of the spin correlations between neighboring sites along the whole system. This is illustrated in Fig. \ref{cargacorrel} for $N=6$ and 7 in the half-filled case. One can notice that the overall antiferromagnetic correlations exhibit a superimposed oscillatory behavior which can be related to the Fermi wavelength of the half-filled case. The suppression of the antiferromagnetic correlations in the superconducting case introduces important differences in the oscillatory pattern in comparison to the normal case.   Due to these different boundary conditions the  patterns have  an opposite phase in one case with respect to the other. On the other hand, there are some differences when comparing even and odd $N$. In the odd case the ideal oscillatory pattern is frustrated and exhibits a jump of $\pi$ in the middle of the chain. 

An interesting effect that appears in the odd $N$ case, when the ground state is doubly degenerate with $S=1/2$, is the non-uniform spatial distribution of the excess spin. This can be appreciated in the inset of Fig. \ref{cargacorrel} where the mean charge per spin along the system is shown. Again the difference between the normal and the superconducting case corresponds to a different boundary condition at both ends of the chain: while in the superconducting case the spin excess exhibits a rather uniform oscillatory pattern along the chain, in the normal case this excess tends to be suppressed at both ends of the chain due to the large antiferromagnetic coupling with the leads. This state is analogous to what the authors of Ref. \cite{bonca}  refer to as two stage Kondo regime in the case of a triple quantum dot system.               

\section{Current phase relation and scaling behavior for large N} 

\begin{figure}[t]
\includegraphics[scale=0.8]{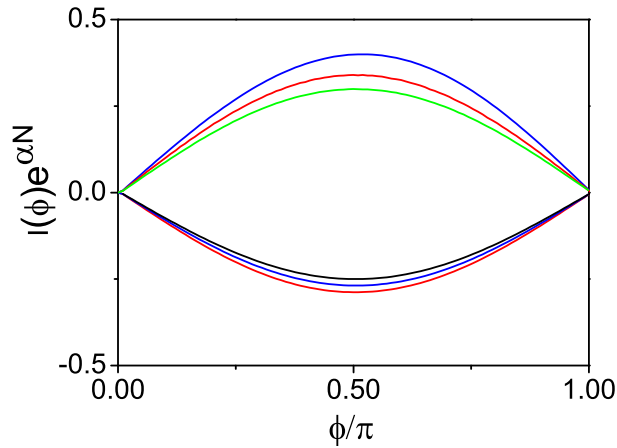}
\caption{(Color online) Current phase characteristics for $U=10$ at half-filling and $N=2,4,6$ (blue, red and green lines) and $N=3,5,7$ (red, blue and black lines). The current is scaled by an exponential $\exp(\alpha N)$ where $\alpha$ has been fitted to a value $\approx 1.8$. $I(\phi)$ is in units of $e\Delta/\hbar$.}
\label{current1}
\end{figure}

In this section we analyze the behavior of the current phase relation for different system sizes. The current phase relation is obtained  from the derivative of the ground state energy with respect to the phase difference $I(\phi)=(e/\hbar)dE(\phi)/d\phi$.
For the sake of simplicity we will concentrate in the half-filled case. 
For the case of repulsive interactions ($U>0$) analyzed so far one finds that typically the current-phase characteristic for sufficiently large $U$ has a sinusoidal form $I(\phi)\approx I_c sin\phi$ with a critical value $I_c$ decreasing exponentially with increasing $N$. This behavior is illustrated in Fig. \ref{current1}. Notice that the results shown in this figure have been scaled with an exponential factor $\exp(\alpha N)$ which nearly merges them into two universal curves one for each $N$ parity.  We have checked that the value of $\alpha$ depends on the interaction strength $U$ roughly linearly. These results are consistent with the prediction of \cite{affleck} stating that the fixed point in a renormalization group analysis for repulsive interactions and in the limit of an infinitely long chain, corresponds to the absence of Josephson coupling. As shown in Fig. \ref{current1} this limiting behavior would be reached in an oscillatory way due to the appearance of $\pi$-junction behavior for odd $N$.
\begin{figure}[h]
\includegraphics[scale=0.8]{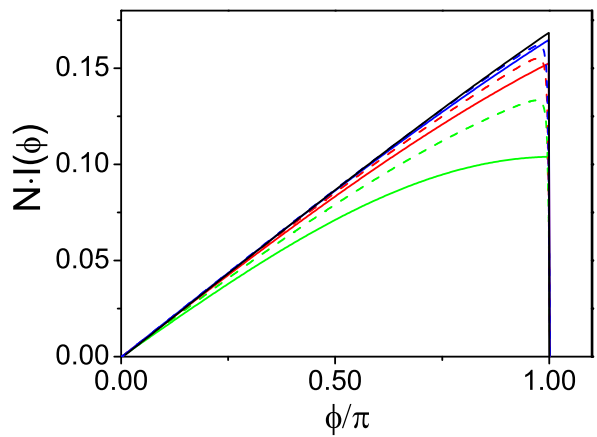}
\caption{(Color online) Current phase characteristics for $U=-10$ at half-filling and $N=1,2,3,4,5,6,7$ from bottom to top. Dashed lines correspond to even $N$. The current is scaled by a factor $N$.}
\label{current2}
\end{figure}

In order to make contact with other predictions of field theoretical calculations for a Luttinger liquid connected to superconducting leads we briefly analyze the behavior of the Josephson current in the presence of an attractive interaction ($U<0$). Let us remark that in this case the only stable ground state corresponds to total spin $S=0$. As shown in Fig. \ref{current2} the current phase relation for increasing $N$ approaches a sawtooth behavior with a critical current scaling as $1/N$. The critical current is also inversely proportional to the absolute value of the local interaction $|U|$. These results are in agreement with the predictions of Ref.\cite{affleck,caux,scalapino}. In particular Affleck {\it et al.} have shown that the renormalization group fixed point corresponds in this case to the condition of perfect Andreev reflection at the interface. Notice that this condition is achieved in the ZBWM  by setting $t=\Delta$ and in the half-filled case.

\section{Concluding remarks}
In the present work we have analyzed the ground state properties of a quantum dot array represented by a finite Hubbard chain  connected to superconducting electrodes. We have shown that the ZBWM  provides a suitable low energy theory for a qualitative description of these systems. This method has allowed us to obtain the phase diagrams in the $\epsilon-U$ plane up to $N\leq10$ sites in the array by numerical diagonalization. 
We have shown that the appearance of $\pi$-junction behavior requires a total odd number of electrons in the array. We have presented an analysis of the spin correlations and their spatial distribution which provides a more detailed insight into the ground state properties. Typically the presence of superconducting electrodes tends to suppress the antiferromagnetic correlations between the chain and the leads. While for the single dot case this suppression is associated with the disappearance of the Kondo effect, for quantum dot arrays the situation is more complex due to the interplay with the internal spin correlations of the chain. A remarkable feature found in this work is the nonuniform spatial distribution of the spin excess characteristic of the $\pi$-junction state. We have finally analyzed the evolution of the current-phase relation and its scaling with increasing $N$. For attractive interactions the results obtained are consistent with previous theoretical studies based on a Luttinger liquid description of the correlated region \cite{affleck,caux}. On the other hand, we have shown that in the half-filled case and for repulsive interactions the critical current decays  exponentially with increasing $N$ although it exhibits an oscillation in the sign due to the alternating occurrence of $0$ and $\pi$ behavior for even and odd $N$.

\section*{Acknowledgments}
The authors would like to thank J. Merino for his valuable help in  the implementation of the Lanczos algorithm and R. Egger for useful discussions and comments.
This work has been financed by the Spanish CYCIT (contract FIS2005-06255).
F.S.B. acknowledges funding by the Ram\'on y Cajal program.

\end{document}